\newcommand{\CLASSINPUTbaselinestretch}{1}
\documentclass[journal]{IEEEtran}
\usepackage{etoolbox}
\makeatletter
\patchcmd{\maketitle}{\@copyrightspace}{}{}{}
\DeclareUnicodeCharacter{2212}{\textendash}
\makeatother
\usepackage{xspace}
\usepackage{siunitx}
\usepackage[]{graphicx}
\usepackage{stfloats}  
\usepackage{cite}  
\usepackage{psfrag}  
\usepackage{subfigure}
\usepackage{wrapfig}
\usepackage{epsfig}
\usepackage{url}
\usepackage{amsmath, amssymb}
\usepackage{array}
\usepackage{multicol}
\usepackage{algorithm}
\usepackage{algorithmic}
\usepackage{mathrsfs}
\usepackage{multirow}
\usepackage[normalem]{ulem}
\usepackage{verbatim}
\usepackage{fixltx2e}
\usepackage[utf8x]{inputenc} 
\usepackage{ucs} 
\usepackage{amsfonts} 
\usepackage{makeidx} 
\usepackage{array} 
\usepackage{colortbl} 
\usepackage{booktabs} 
\usepackage[justification=centering]{caption}
\usepackage[table]{xcolor}

\usepackage{pifont}
\definecolor{tableheading}{rgb}{0.9,0.9,0.9}
\definecolor{softblue}{rgb}{0.8,0.8,1} 
\usepackage{tabu}
\usepackage{floatflt}
\usepackage{amsmath, amssymb}
\usepackage{tikz}
\usepackage{multicol}
\usepackage{xcolor}
\usepackage{amsmath,amssymb}
\usepackage{tabularx}
\usepackage{color}
\usepackage[dvipsnames,prologue]{pstricks}

\begin{document}

\title{
  Evaluation of STT-MRAM as a Scratchpad for Training in ML Accelerators}
\author{\IEEEauthorblockN{Sourjya Roy\IEEEauthorrefmark{1}, Cheng Wang\IEEEauthorrefmark{2}, and Anand Raghunathan\IEEEauthorrefmark{1}\\}
\IEEEauthorblockA{\IEEEauthorrefmark{1}School of Electrical and Computer Engineering, Purdue University, West Lafayette, IN, USA\\
\IEEEauthorrefmark{2}Department of Electrical and Computer Engineering, Iowa State University, Ames, IA, USA}\\
{\tt roy48@purdue.edu, chengw@iastate.edu, raghunathan@purdue.edu
}}
\maketitle



\sloppy

\maketitle

\begin{abstract}
Progress in artificial intelligence and machine learning over the past decade has been driven by the ability to train larger deep neural networks (DNNs) with increased amounts of data, leading to a compute demand that far exceeds the growth in hardware performance afforded by Moore's law. Training DNNs is an extremely memory-intensive process, requiring not just the model weights but also activations and gradients for an entire minibatch to be stored. Inability to store these data structures on chip results in expensive off-chip memory accesses, limiting both training speed and energy efficiency. Current training accelerators use large SRAM-based scratchpads, but the need to provide larger on-chip memory capacities and mitigate the leakage power of SRAM motivates the exploration of more dense non-volatile memory technologies. Among the emerging non-volatile memory (NVM) options, we observe that spintronic memories like Spin-Transfer-Torque MRAM (STT-MRAM) offer several desirable properties for training accelerators, including high endurance and reasonable access time. Compared to SRAM, STT-MRAM provides 3-4 times higher density and significantly reduced leakage power on the one hand, while requiring higher write energy and write time on the other hand. The inefficient writes are due to the need for a higher write voltage and longer write duration to ensure reliable switching. 

In this study, we perform a comprehensive device-to-system evaluation and co-optimization of STT-MRAM for efficient ML training accelerator design. We devised a cross-layer simulation framework to evaluate the effectiveness of STT-MRAM as a scratchpad in a systolic-array-based DNN accelerator. We consider two design scenarios --- replacing SRAM scratchpads with iso-capacity and iso-area equivalent versions of STT-MRAM scratchpads. We further propose to address the inefficiency of writes in STT-MRAM by utilizing reduced write voltage and duration, which come at the cost of introducing errors. This is based on the observation that the intrinsically error-resilient nature of DNN training can tolerate such errors to an extent. To evaluate the ensuing accuracy-efficiency trade-off, we conduct a thorough analysis of the error tolerance of input activations, weights  and errors during the training process. We propose heterogeneous memory configurations that enable training convergence with good accuracy. Our results indicate that replacing SRAM with STT-MRAM can provide up to 15x and 22x improvement in system level energy for iso-capacity and iso-area scenarios respectively, across a suite of DNN benchmarks (AlexNet, VGG16, ResNet18 and ResNet34). Further optimizing STT-MRAM write operations can provide over 2x improvement in write energy for minimal degradation in application level training accuracy.

\end{abstract}

\section{Introduction}
\label{sec:introduction}
{\noindent} 
\vspace*{-0pt}
Modern machine learning (ML) algorithms, particularly deep neural networks (DNNs), have achieved remarkable performance, surpassing human capabilities across diverse tasks in computer vision and language processing~\cite{roser_2022}. However, this success has been accompanied by a rapid growth in the computational requirements for DNN processing, resulting in great interest in efficient hardware accelerators specifically designed for DNNs. A predominant factor contributing to the computational challenge of DNN workloads is their memory-intensive nature \cite{ibm_blog}; consequently, a substantial volume of data must be transferred from off-chip main memory to on-chip processing elements. This data movement gives rise to a critical bottleneck in hardware performance, commonly referred to as the von-Neumann or memory wall bottleneck. Larger on-chip memory can potentially mitigate this bottleneck by storing more data and feeding more accesses without going off-chip. Indeed, this has been the driving factor behind the use of ever-increasing scratchpads in DNN accelerators~\cite{sze_dnn_hw_overview,ibm_scratchpad}. Nevertheless, on-chip memory built on standard static random-access-memory (SRAM) occupies considerable space, requiring a minimum of six transistors (6T) per cell. Additionally, the scaling of SRAM cells at advanced nodes (7nm and beyond) has considerably slowed down \cite{tsmc_sram_2022} \cite{wiki_chip}, resulting in diminishing gains in effective memory density in terms of bits per unit area. The net effect of these trends is evident in today's DNN accelerators where a majority of the chip area is consumed by  scratchpads~\cite{ten_lessons_google,sze_dnn_hw_overview}. 

An alternative pathway to enhancing on-chip memory capacity involves leveraging emerging embedded non-volatile memories (NVMs), whose compact cell sizes contribute to improved on-chip scratchpad capacity. Additionally, the non-volatile nature of NVMs offers substantial reductions in leakage power when compared to CMOS-based memory technologies. The combination of compactness and reduced power consumption positions embedded NVMs as a highly promising solution for overcoming the memory challenges associated with DNN accelerators. We note that the use of NVMs for on-chip storage also opens up the possibility for in-memory computing~\cite{ibm_analogai_overview,purdue_xbar_overview}, leading to further potential improvements in energy efficiency. While this approach is promising and has been an acrive area of research over the past decade, it nevertheless faces considerable challenges related to analog noise and non-idealities, overheads of peripheral circuits, {\em etc.} that are being actively addressed. Notwithstanding the potential of STT-MRAM for in-memory computing~\cite{spindle}, we focus on the nearer-term use case for NVMs as scratchpad replacement since it is not subject to the above challenges.

Most efforts exploring NVMs as scratchpads in DNN accelerators focus on inference, where NVMs are used to store the pre-trained DNN model weights. While inference is an important type of workload, it is also important to address the significantly larger computational demands of training DNNs. Training is a highly memory intensive process due to the need to store several data structures like weights, minibatch activations and error and weight gradients. It is also important to note that training performs large numbers of writes to the memory storing these data structures, demanding high device endurance. The adoption of embedded NVMs for training remains largely hindered by the limited endurance as well as costly write operations of NVMs compared to SRAMs. Specifically, the leading NVM candidates, such as resistive memory (ReRAM) and phase change memory (PCM) require sophisticated pulsing schemes ($>$ 10\textsuperscript{2} ns) to set/reset during write operations and suffer from poor write endurance \cite{pcm_pulsing}\cite{pcm}\cite{reram}. In contrast, STT-MRAM provides the highest endurance ($>$ 10\textsuperscript{15} cycles) with 1 to 10 ns access time \cite{mram_endurance} while also providing 3-4X higher density compared to SRAM. Therefore, STT-MRAM can be considered the most promising NVM technology for DNN training accelerators.

Notwithstanding its advantages, when compared to SRAM at the same technology node, standard STT-MRAM still exhibits higher write energy and latency. This is primarily attributed to the need for large write voltages and long write pulse durations to ensure reliable switching of the magnetization. 
The high writing cost of STT-MRAM presents a major barrier that needs to be addressed to fully leverage its potential for use in DNN training accelerators.

In this work, we investigate device-to-system co-optimization of STT-MRAM for efficient DNN training acceleration. 
We propose to exploit MRAM with reduced write voltage and write duration to address the large cost of write operations, and evaluate the impact of the low-cost write operations on both energy-to-train and the accuracy of the resulting DNN models. 
We systematically analyze the error tolerance of input activations and weights during training, and propose a heterogeneous memory architecture wherein different parts of a number (e.g., mantissa and exponent) are mapped onto memory arrays with different write error rates.

In summary, this work makes the following contributions:
\begin{enumerate}
    \item We develop a cross-layer design space exploration framework to evaluate STTMRAM as a scratchpad in DNN training accelerators and compare it with SRAM under iso-capacity and iso-area scenarios to study the potential energy benefits. 
    \item We explore the use of reduced write voltage and duration to address the STT-MRAM write bottleneck.
    Technology-dependent bit errors are incorporated into the training of DNNs for standard image classification tasks to study the impact of write errors on application-level accuracy. To mitigate this impact, particularly in the low-energy write regime, we observe that the different parts of a word (sign, exponent and mantissa) have distinct error tolerances. Based on this insight, we
    propose a heterogeneous memory organization where these parts are mapped to different STT-MRAM arrays that are operated with different write voltages and durations.  
    \item Replacing STT-MRAM with SRAM as an on-chip scratchpad, we achieve up to 15x and up to 23x improvement in energy efficiency for iso capacity and iso area scenarios, respectively. Further optimization of STTMRAM for writes can provide more than 2x improvement in system level write energy at the cost of minimal accuracy loss at application level.
    
\end{enumerate}

\vspace*{2pt}
\section{Related Work}
\label{sec:relatedWork}
NVM technologies have been explored to provide high-density and low-power embedded memory for efficient ML hardware systems. In particular, MRAM has emerged as a leading candidate thanks to its demonstrated scalability, high energy efficiency (compared to other NVMs), and superior endurance. In general, the explorations of exploiting MRAM for ML acceleration can be grouped into two categories: MRAM for digital hardware accelerators and MRAM for analog and mixed-signal in-memory computing. 

Several recent works have explored using MRAM for digital accelerator architectures in various places including caches, global buffer, and 3D-stacked high bandwidth memory. 
DeepNVM \cite{DeepNVM} introduced a technology-aware simulation framework that demonstrate performance improvement when MRAM is implemented for last-level caches in GPU architectures for executing deep learning workloads. 
Authors in \cite{mram_asic} presented a domain-specific convolutional neural network (CNN) accelerator using co-designed STT-MRAM for on-chip neural network weight storage, and demonstrated energy-efficient ML inference. 
A hierarchical hybrid embedded memory system for systolic array was designed to thoughtfully partition the mapping of different layers in a CNN model \cite{mram_hbm} into 3D-stacked STT-MRAM and on-die SRAM. Such memory systems could exploit the non-volatility and high density of MRAM for storing pre-trained models, and at the same time leverage the low latency on-die SRAM to enable efficient real-time on-chip learning. 
Another recent work looked into directly using MRAM as the on-chip buffer in a DNN inference accelerator\cite{mram_eb}. In order to maximize the density of on-chip MRAM buffer and thus minimize the off-chip DRAM access latency and energy, \cite{mram_eb} proposed to relax the thermal stability of MRAM cell in order to achieve efficient and fast write operations with reduced memory cell area. 
While such MRAM design with reduced retention exhibits satisfactory inference accuracy, it is unclear whether the MRAM with elevated bit errors due to the reduced thermal stability would still function for ML training. Moreover, the scaled thermal stability approach can only be done at fabrication, lacking the flexibility required for processing workloads in various application scenarios. In contrast, our presented work focuses on accelerating ML training with customized MRAM design. The proposed MRAM-based scratchpad design with bit-level granularity and reconfigurable heterogeneity offers the flexibility to handle diverse ML workloads with distinct error resilience while achieving improved hardware efficiency for DNN training.  

In addition to the exploration for digital accelerators, a plethora of works have also investigated developing in-memory computing architectures with MRAM arrays\cite{ibm_analogai_overview,purdue_xbar_overview,spindle,samsung_xb, naresh_mram, logic_in_memory_deliang, xbar_opt, sot_imec}. While in-memory computing with MRAM for non-volatile weight storage demonstrates great potential for efficient ML inference, the adoption of in-memory computing at large scale still faces immense challenges such as device/circuit non-idealities, limited sensing margin, and large overhead of peripheral circuits. Addressing these challenges remains an active area of research.
 
\vspace*{6pt}
\section{Preliminaries}
\label{sec:preliminaries}
\subsection{DNN Training}
\begin{figure}[htb]
    \centering
    \vspace*{-2pt} 

\includegraphics[width=\columnwidth]{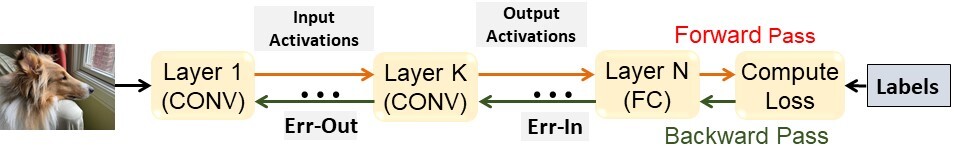}
    \caption{DNN Training}
	\vspace*{-2pt}
    \label{fig:DNN_train}
\end{figure}
DNN (Deep Neural Network) training, depicted in Figure \ref{fig:DNN_train}, is a learning process aimed at adjusting the parameters (weights) of a DNN to maximize its performance on samples in the training dataset. The model's parameters are initialized randomly and then iteratively updated during the training process, guided by a loss function and an optimizer. The learning rate, a hyperparameter, determines the speed of learning by modulating the magnitude of the parameter updates. The training process involves three stages: Forward Propagation, Back Propagation, and Weight Update, which we describe in turn below.
\\
\noindent\textbf{Forward Propagation:} During the process of forward propagation, a mini-batch of input activations belonging to a specific layer is multiplied by the corresponding weights or parameters associated with that layer. This multiplication yields a mini-batch of output activations, which subsequently serve as the input for the subsequent layer. The propagation of these activations continues until they reach the final layer, where an output prediction is generated. The data structures necessary for  this phase are the activations and weights.
\\
\noindent\textbf{Back Propagation:} The final layer plays a pivotal role in computing the error by comparing the predicted output with the true output using the selected loss function. In the subsequent phase known as Back Propagation, this error is propagated in a reverse manner, starting from the final layer and reaching all the way back to the first layer. At each layer, the gradients of the error with respect to the weights and input activations are computed. To facilitate the execution of this phase, the essential data structures include the activations, weights, and error gradients. These data structures are crucial for accurately adjusting the network's parameters and fine-tuning the model based on the computed gradients.
\\
\noindent\textbf{Weight Update:} During the learning process, the error gradients are crucial in updating the weights. These gradients, which represent the changes needed in the weights, are multiplied by the learning rate before being utilized to update the previous weight values. This update operation ensures that the new weights obtained will be used in the subsequent forward pass. It is important to note that the sign of the gradients determines whether the new weight value will be higher or lower than the previous one. To facilitate this phase, two key data structures are necessary: the weights and the weight gradients.
\vspace{-3mm}
\subsection{Embedded STT-MRAM}
\begin{figure}[H]
    \centering
    \vspace*{-2pt} 

\includegraphics[width=0.9\columnwidth]{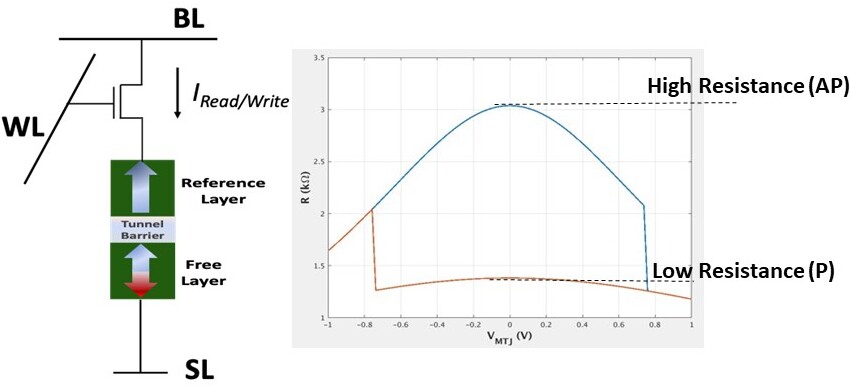}
    \caption{STTMRAM}
	\vspace*{-2pt}
    \label{fig:STTMRAM_train}
\end{figure}
The STT-MRAM cell is based on a magnetic tunnel junction (MTJ) in series with its access transistor (Figure \ref{fig:STTMRAM_train}). An MTJ comprises of an insulating tunnel barrier (commonly made of MgO) sandwiched between two ferromagnetic layers, i.e., a fixed reference layer
(RL) and a free layer (FL). The MTJ provides bi-stable resistance states that follow the relative orientation of the two layers being parallel (P) or anti-parallel (AP). The tunneling
magneto-resistance ratio (TMR = (R\textsubscript{AP} $-$ R\textsubscript{P})/R\textsubscript{P}) reflects the device OFF/ON ratio, which impacts the sensing margin.
The magnetization of FL is switched by STT from the spin-polarized current passing through the MTJ due to the interaction of local magnetic moments and electron spins in the conduction current. 
In order to provide thermal stability and minimize errors due to thermal agitation and process variation, the thermal stability factor K\textsuperscript{u}V/k\textsubscript{B}T needs to be about 45-75. The design choice of thermal stability depends on the system-level requirement for data retention. Since the uniaxial anisotropy constant K\textsuperscript{u} is a material-dependent constant, the variation of thermal stability factor is usually achieved by modifying the volume of the FL in an MTJ.
MTJs with perpendicular anisotropy  offer the potential of achieving higher density and more efficient switching dynamics. 

Compared to SRAM which normally requires at least 6 transistors per bit-cell, the 1T-1MTJ STT-MRAM provides 2-4x higher density with substantially reduced leakage power. At the cell level, although STT-MRAM is already improved in write efficency compared to field-driven switching, the write energy and latency of STT-MRAM is about 10x higher than SRAM. Due to the large write current density required, the MRAM cell area is determined by the access transistor (not the MTJ pillar size). Today, novel technologies such as spin-orbit torque (SOT) and voltage-controlled magnetic anisotropy (VCMA) are being explored to further improve the MRAM write performance, although these technologies are still in their nascent stage of development.

\subsection{Baseline accelerator architecture and dataflow}
\begin{figure}[H]
    \centering
    \vspace*{-2pt} 

\includegraphics[width=0.9\columnwidth]{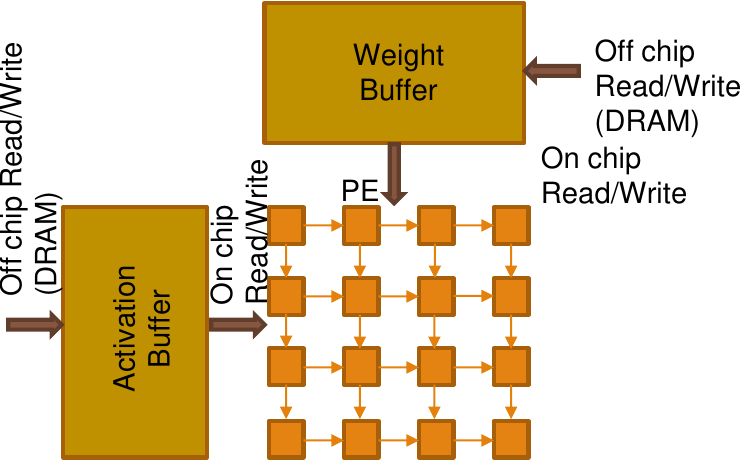}
    \caption{Systolic array accelerator with activation and weight buffers}
	\vspace*{-2pt}
    \label{fig:systolicarray}
\end{figure}
Figure \ref{fig:systolicarray} presents the architecture of a simple systolic-array-based DNN accelerator used in our evaluation. A systolic array comprises a 2D array of processing elements (PEs), where each PE is interconnected with all or a subset of its nearest neighbor PEs. When mapping DNN computations to systolic arrays, it is important to choose a suitable dataflow~\cite{sze_dnn_hw_overview}. In the output stationary dataflow, each PE in the systolic array stores an output while the computation of a DNN layer is being performed. Inputs and weights are streamed into the systolic array from different directions at each cycle, enabling in-place reduction at each PE. Let us consider a forward pass in a convolution layer as an example. An input layer with dimensions of B*I*M*N undergoes convolution with a weight layer of dimensions O*I*K*K. Here M and N refers to the input height and width, B refers to the batch size, K refers to the kernel width and I and O refer to the number of input and output channels, respectively. K\textsuperscript{2}I scalar values are streamed in from activation and weight buffers and reduced at every PE to produce one output activation value. 
The dimensions of the systolic array determine the level of temporal reuse within it.
\vspace*{-0pt}



\vspace*{2pt}
\section{Experimental Methodology}
\label{sec:exptsetup}
\vspace*{0pt}
\begin{figure}[htb]
    \centering
    \vspace*{-2pt} 

\includegraphics[width=0.9\columnwidth]{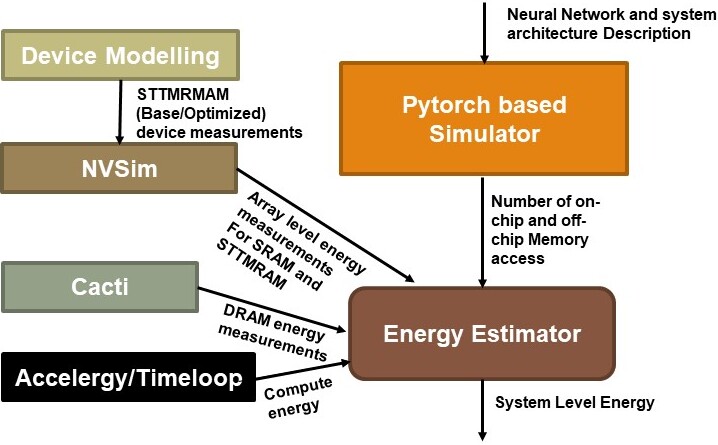}
    \caption{Cross Layer Evaluation Framework}
	\vspace*{-2pt}
    \label{fig:framework}
\end{figure}
To enable the evaluation of STT-MRAM based scratchpads in DNN training accelerators, we developed the cross-layer framework illustrated in Figure \ref{fig:framework}. The framework includes energy simulation of a hardware system comprised of on-chip memory, a compute unit based on systolic array architecture, and off-chip DRAM. The device-to-array simulation of on-chip memory includes STT-MRAM modeling using a MATLAB-based macro-spin simulator for capturing the magnetic switching dynamics. The device-level behaviors obtained from MATLAB are provided to NVSim \cite{nvsim}, a simulation tool for evaluating the performance, energy, and area of NVM-based on-chip memory. NVSim provides latency, area, and energy estimates for memory arrays of different capacities based on the specifications of memory cells. The results of both MRAM and the baseline SRAM arrays in our experiments are obtained using NVSim. The technology-dependent specifications of on-chip memories, namely STT-MRAM and SRAM, are based on a 28nm technology node. The energy estimation for off-chip DRAM is performed using the CACTI tool \cite{cacti6}. DRAM is modeled using the DDR3 interface and a 65nm technology node. 
The evaluated compute unit is based on a systolic array architecture with output stationary data flow. The compute energy for the systolic array is obtained using Accelergy \cite{accelergy}. The system-level evaluation considers a systolic array size of 256x256 with an on-chip clock frequency of 1 GHz. 

To evaluate the system-level energy efficiency, we developed a simulator based on PyTorch. This simulator takes both algorithmic attributes (such as the DNN architecture) and hardware attributes (such as the accelerator architecture) as inputs. The algorithmic attributes of DNN workloads include the layer type (convolution and fully connected), dataset, and layer dimensions, while the hardware attributes include the data flow type, systolic array size, and on-chip memory sizes. Based on these input parameters, the PyTorch simulator calculates the number of on-chip memory accesses, off-chip DRAM accesses, and leakage cycles. The estimations from NVSim, CACTI, and Accelergy, along with the output from the PyTorch-based simulator, are sent to an energy estimator to calculate the total output energy. 

We make the following assumptions for the system-level evaluation:
\begin{enumerate}
\item Three types of buffers are considered in the evaluated architecture: activation buffer, weight buffer, and error buffer. However, there is no separate output buffer included due to the output stationary dataflow. The total on-chip memory capacity is divided equally among the activation buffer, weight buffer, and error buffer.
\item Accesses to DRAM and the on-chip memory array are executed in a sequential manner, one after the other, rather than concurrently or in a parallel fashion.
\item 
The values of a data structure, such as activations, weights, and errors, are stored in the on-chip memory array only if the entire data structure for a layer can fit within the available on-chip memory capacity. If the size of a data structure exceeds the capacity of the on-chip memory array, it would be stored in off-chip memory instead. This is typical of software-managed scratchpads where each data structure is treated as atomic.
\item 
During the forward pass of the neural network, data structures (such as activations, weights, and errors) are removed from the memory array in a serial manner. This means that data structures are evicted from the memory array one by one, typically in the order they were stored, to make space for incoming data structures. Similarly, during the backward pass of the neural network, data structures are also removed from the memory array in a serial manner, but in the reverse order compared to the forward pass. This ensures that the data structures required for the backward pass are evicted in the correct order, allowing for proper gradient calculations and parameter updates. 
\end{enumerate}

\section{Results and Discussions}
In this section, we will discuss the results from the proposed cross-layer exploration. First, we will delve into the probabilistic switching in STT-MRAM write operations. We focus on simulating the STT-driven switching dynamics of MTJ and evaluating the elevated write error rates due to probabilistic switching under write voltage with reduced magnitude and/or reduced duration. 
Next, at the array level, comparison between SRAM and STT-MRAM will be performed under iso-capacity and iso-area scenarios when STT-MRAM designs with both standard and low-overhead write operations will be evaluated.
Subsequently, we will evaluate the system-level energy consumption of running DNN training workload in a ML accelerator with systolic array architecture. Both SRAM and STT-MRAM will be designed as the scratchpad in systolic array architecture with an output stationary dataflow. We will compare the energy efficiency of the hardware using MRAM and SRAM under iso-capacity and iso-area scenarios.
Lastly, 
we will investigate the impact of write optimization on application-level accuracy and the trade-off between hardware efficiency and algorithmic accuracy. After incorporating the elevated write errors into our PyTorch-based simulator, we propose a hybrid architecture with distinct memory write settings for different bits of the floating point representation. The proposed hybrid memory architecture enables energy-efficient MRAM writing, leading to further improvement in the write energy of memory arrays while ensuring minimal degradation of DNN accuracy. 

\begin{figure}[H]
  \centering
  \includegraphics[width=0.9\columnwidth]{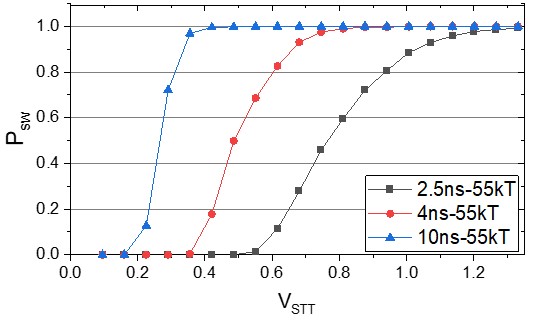}
  \caption{Switching probability of MRAM cell under varying write voltage durations and magnitudes.}
  \vspace{-4mm}
  \label{fig:psw}
\end{figure}
\subsection{Modeling of the STT-MRAM write operations}
\subsubsection{Simulation of MTJ switching dynamics}

We investigate the write operations of STT-MRAM cell based on macro-spin simulations of the magnetization switching. The magnetization dynamics of the FL of MRAM cell can be modeled by solving the Landau-Liftshitz Gilbert equation with the STT term incorporated. Based on the recently reported STT-MRAM device fabrications, the MTJ for 28nm STT-MRAM is set to have a lateral dimension 35-40 nm in our simulations, and the free layer CoFeB has interfacial perpendicular magnetic anisotropy. 
The detailed simulation parameters are summarized in Table \ref{tab:mtj-param}. 
\begin{table}
\small
\centering
\begin{tabular}
{|c|c|}
\hline
\textbf{Parameter} & \textbf{Value}  \\
\hline
FL thickness t\textsubscript{CoFeB} & 1.0 nm\\
\hline
MTJ lateral dimension  & 35nm x 35nm\\
\hline
Gilbert damping $\alpha$ & 0.006 \\
\hline
Anisotropy constant \textit{K} &  1.0 ergs/cm\textsuperscript{2}\\
\hline
Saturation magnetization & 1200 emu/cc\\
\hline
\end{tabular}
\caption{\label{tab:mtj-param} Key parameters of device-level macro-spin simulation of the STT-driven switching.}
\end{table}

Since the onset of STT switching is attributed to the combination of the thermal agitation at room temperature and the directional torque exerted from the currents, there is intrinsic stochasticity in MRAM switching. We model the stochasticity originated from thermal perturbation as an effective perturbation field \cite{thermal_field}:
\begin{equation} \label{thermal}
H_{Thermal} = \mathcal{N} (0,1)\sqrt{\frac{2 \alpha k_{B}T}{\gamma M_{S}V\delta t}},
\end{equation}
where $\mathcal{N}$(0,1) is a Gaussian random variable with mean and stand deviation being 0 and 1 respectively, and $\delta$t is the simulation time step. 
As a result of the stochastic switching, varying probabilities $P_{sw}$ can be observed under applied write voltages with different magnitudes and durations. As is illustrated in Figure \ref{fig:psw}, the switching probability $P_{sw}$ exhibit a sigmoid-like behavior as the excitation current increases. The onset of switching occurs at smaller currents when the excitation duration is longer.
We obtain that the STT efficiency $\eta = 1.15$   
 k\textsubscript{B}T/$\mu A$ ($\eta$ defined as the ratio of energy barrier and critical current $\eta = E\textsubscript{B}/I\textsubscript{c0} $), which is consistent with the reported values of 1.0$-$3.0 in recent fabrication works \cite{ibm_mram_param} \cite{tdk_mram} . It has been demonstrated that reducing the energy barrier E\textsubscript{B} by reducing the MTJ dimension will lead to a trade-off between thermal stability and lower write current \cite{ibm_mram_param} \cite{toshiba}. However, it is important to note that lowering the energy barrier will make the MTJ susceptible to process variations and thermal noises, and modifying the MTJ device dimension post fabrication is not feasible. Therefore, we chose to explore our MRAM design with a sufficient energy barrier E\textsubscript{b} = 55 k\textsubscript{B}T that ensures stable data retention at the room temperature for several months.

\subsubsection{Write error rate (WER) with reduced magnitude and duration of STT voltage}
\begin{figure}[htb]
  \centering
  \includegraphics[width=0.8\columnwidth]{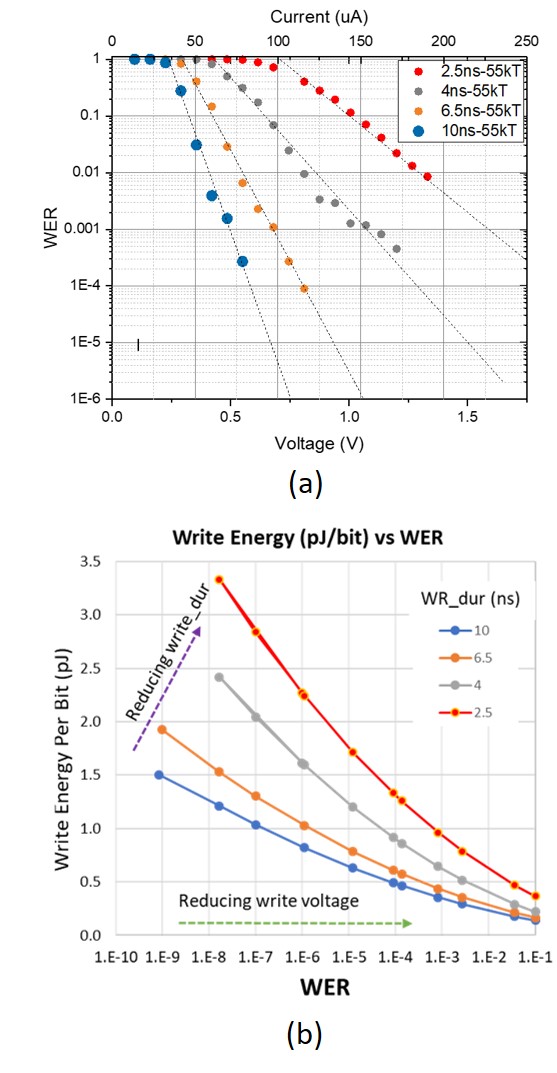}
  \caption{(a) WER of MRAM under varying write voltage durations and magnitudes. (b) Trade-off of write energy and WER based on the macrospin simulations.}
  \vspace{-4mm}
  \label{fig:WER}
\end{figure}
The switching probability $P_{sw}$ is linked to the write error rate (WER) of MRAM:

\begin{equation} \label{eq_WER}
ln WER = ln\left(1-P_{sw}\left( V_{STT}, t_{STT}\right)\right), 
\end{equation}
where $V_{STT}$, $t_{STT}$ are the magnitude and duration of the applied write voltage, respectively. 

Figure \ref{fig:WER} (a) shows the WER as a function of write current/voltage under varying write pulse durations. It is observed that the WER remains around 1.0 (no switching) until the write current reaches a threshold. After the onset of switching, WER decreases exponentially with the increasing current, as suggested by the approximately constant slope in such ln(WER) plot. Such observation of constant slope in ln(WER) plots is consistent with the Fokker-Planck theory based on macro-spin approximation \cite{}. In our experiment, we run magnetic switching dynamics for 20K iterations to capture the WER down to ln(WER)$\sim$-4. We obtain the required write currents for WER valued from 1E-5 to 1E-9 based on linear extrapolation of the \textit{ln(WER)} vs. V\textsubscript{STT} curves.

The design of robust on-chip MRAM requires to reach WER $<$ 1E-9 in order to inhibit bit errors during memory updates. Therefore, we set the WER $\sim$ 1E-9 (8.62E-10) as our baseline for setting the write current's magnitude and duration. As shown in Figure \ref{fig:WER} (b), lowering $V_{STT}$ will lead to reduction of write energy per bit at the cost of deteriorating the WER. Moreover, while reducing the $t_{STT}$ leads to latency improvement, ensuring a certain WER requires more write energy per bit. This is due to the requirement of applying larger STT currents when the write duration is reduced.
Next, we will incorporate the modifications of $V_{STT}$ and $t_{STT}$ in MRAM cell into memory array and micro-architecture design, and elucidate the implications of
device cell modifications on the hardware performance at the micro-architecture level.

\subsection{Array-level design of on-chip memory}
\begin{figure*}[!t]
  \centering
  \includegraphics[width=0.9\textwidth]{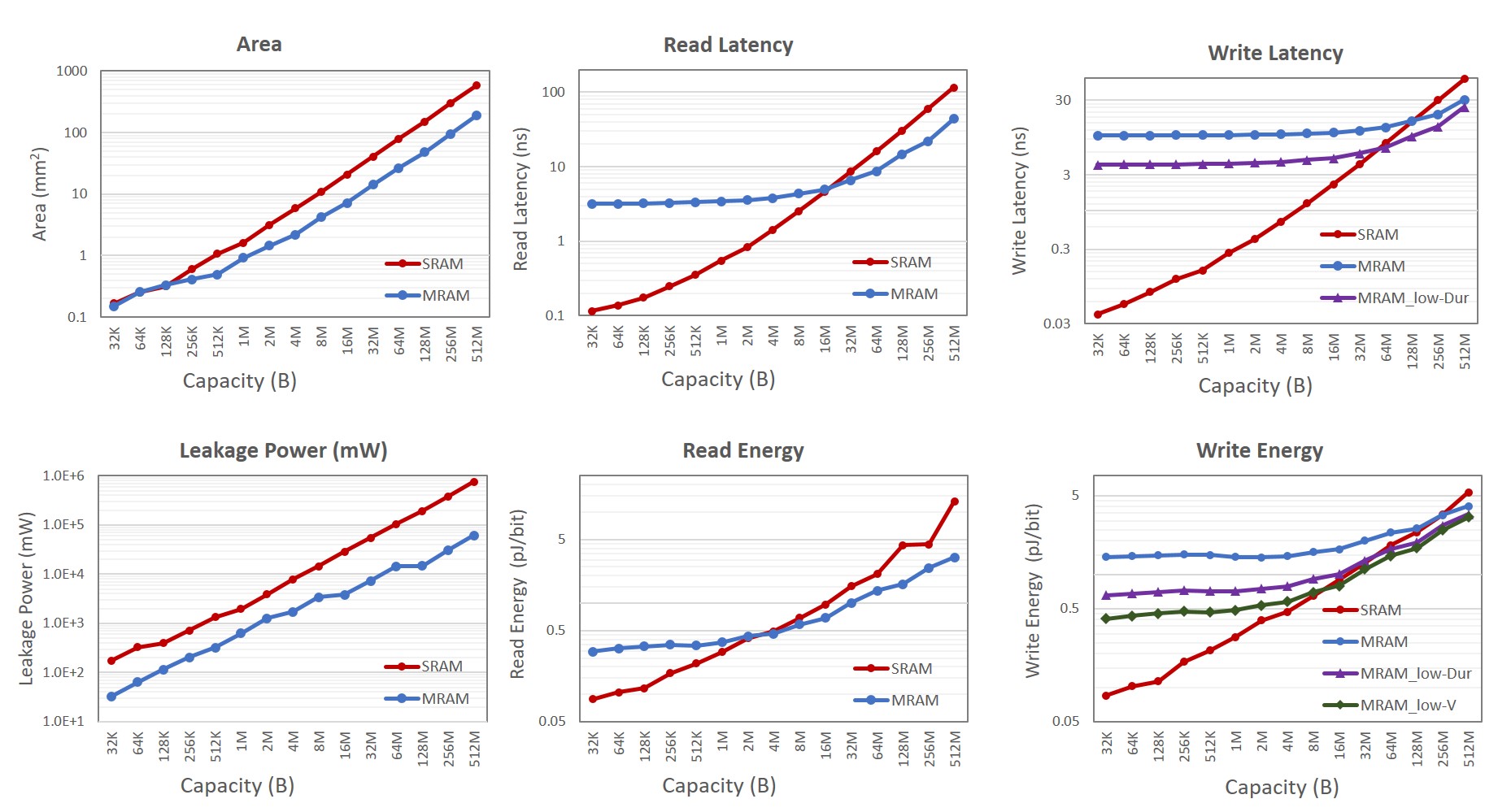}
  \caption{Iso-capacity analysis of SRAM and STT-MRAM memory design based on NVsim simulations for capacity 32 KB to 512 MB. For MRAM with reduced write duration (low-dur) and reduced write voltage (low-V), the WER is elevated to 8E-4.}
  \vspace{-4mm}
  \label{fig:nvsim}
\end{figure*}
In this sub section, we evaluate the array-level hardware performance of STT-MRAM in comparison with the SRAM. We employ NVsim \cite{nvsim} to analyze the circuit-level energy efficiency, latency, and area based on varying cell designs including standard SRAM, standard MRAM, and MRAM with low-cost write operations. 
The simulated MRAM bit cell parameters are summarized in \textbf{Table 2} together with SRAM baseline. 
Note that the NVsim simulator contains various optimization targets that may lead to variance in the simulation results. After experimenting various optimization targets, we select \textit{cache} configuration with \textit{Read Latency} as the optimization target, which works well for the whole range of memory capacity in our exploration.

We first provide the iso-capacity analysis of circuit-level simulations for capacity ranged from 32 KB to 512 MB. In general, the the capacity increases, both SRAM and STT-MRAM consume larger area and increased leakage power due to increased numbers of transistors employed. As a result of larger arrays and larger banks, the parasitic resistance and capacitors in the circuit also increases, leading to increased latency and read/write energy.

WE further observe that compared to SRAM, STT-MRAM starts to exhibit advantages in area efficiency as the memory capacity exceeds 128 KB. Thanks to the compact cell size, MRAM achieves 2-3x better area efficiency for the wide range of capacity from 1MB to 512 MB. Moreover, the leakage power of STT-MRAM is substantially reduced compared to SRAM thanks to the non-volatility of magnetic storage. We further observe that while the read/write latency and energy of SRAM is significantly lower than MRAM at the cell level, the array-level latency and energy of SRAM will increase with the capacity at a much faster rate compared to MRAM. Such significant increased cost at the array-level is due to the significantly larger SRAM cell area. As capacity increases, SRAM arrays consume much larger areas and longer wires along the bitlines and wordlines, leading to more severe parasitic resistance and capacitance that dominate the access latency and energy at higher capacity.

We further explore the array-level MRAM performance with reduced write overheads. Both reduced write voltage and shortened write duration are evaluated. As is shown in Figure \ref{fig:nvsim}, we observe that with reduced write voltage and write latency, the crossover between STT-MRAM and SRAM in their array-level metrics will shift to smaller capacity, indicating more performance advantage from MRAM. Such improvement in hardware performance is achieved with the trade-off of elevated WER due to the induced non-deterministic switching. For example, compared to the baseline MRAM, the memory design with reduced write overhead can lead to 53$\%$ reduction in latency and 60$\%$ access energy at a cost of increasing the WER to 0.8E-4. We will exploit the MRAM design optimization with low write costs and investigate the impacts of the inflated WER for DNN training.
\begin{figure}[!t]
  \centering
  \includegraphics[width=0.9\columnwidth]{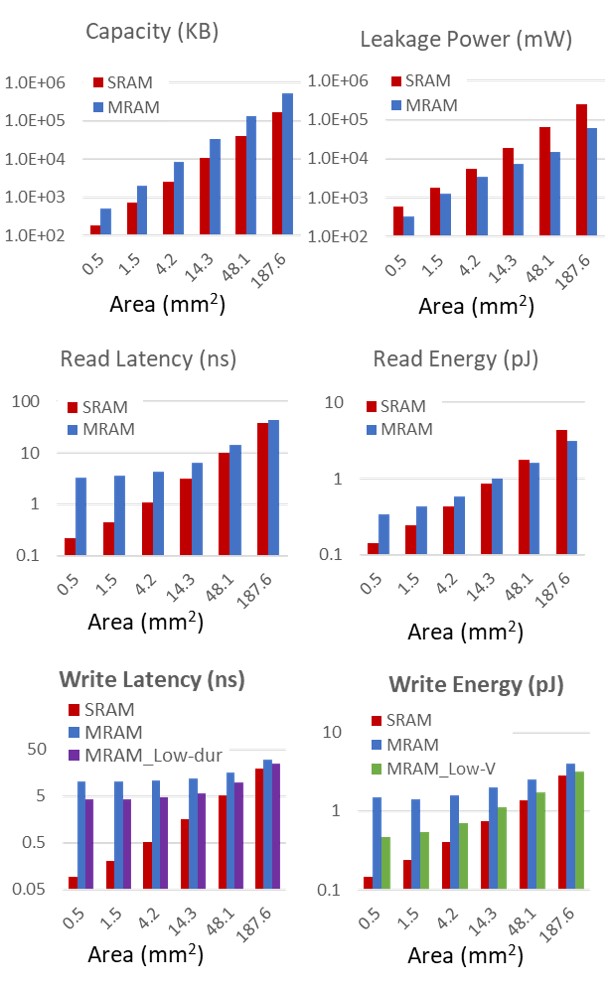}
  \caption{Iso-area analysis of SRAM and STT-MRAM design for area budgets from 0.5 to 187 mm\textsuperscript{2}. For MRAM with reduced write duration (low-dur) and reduced write voltage (low-V), the WER is elevated to 8E-4.}
  \vspace{-4mm}
  \label{fig:iso-area}
\end{figure}

\begin{table}
\small
\begin{tabular}
{|l|m{0.8cm}|m{0.8cm}|m{0.8cm}|m{0.8cm}|}
\hline
Area (mm\textsuperscript{2})& \multicolumn{2}{|c|}{0.5} & \multicolumn{2}{|c|}{48.1} \\
\hline
Memory type &SRAM &MRAM &SRAM &MRAM\\
\hline
Capacity (KB) &183 &512 &40592 & 131072\\
\hline
Read latency (ns) & 0.2 & 3.3 & 10.3 & 14.6\\
\hline
 Write latency (ns) & 0.1 & 10.2 & 5.3 & 15.8 \\
\hline
Read energy (pJ) & 0.1 & 0.3 & 1.8 & 1.6 \\
\hline
Write energy (pJ) & 0.1 & 1.5 & 1.4 & 2.6\\
\hline
Leakage Power (mW) & 594 & 323 & 64257 & 14573\\
\hline
\end{tabular}
\caption{\label{tab:iso-area} Iso-area comparison of SRAM and baseline STT-MRAM under small (0.5 mm\textsuperscript{2}) and large (48.1 mm\textsuperscript{2}) area budgets.}
\end{table}


We also present an iso-area analysis so that the comparison of MRAM and SRAM could be drawn in more practical scenarios. As summarized in Figure \ref{fig:iso-area}, hardware cost of both SRAM and MRAM increase at larger area.
While SRAM overall has lower access latency and energy, the performance gap between SRAM and MRAM under iso-area shrinks rapidly as the chip capacity grows under larger area. 
Such observations reflect the increasing impact of circuit parasitics on the array-level performance as the chip area grows, which is consistent with the observation from the iso-capacity curves in Figure \ref{fig:nvsim}. Moreover, MRAM designs with reduced write duration and reduced write voltage exhibit further narrowing of the performance gap between MRAM and SRAM. 
We summarize in Table \ref{tab:iso-area} the array-level memory performance of SRAM and MRAM under two representative area budgets. We observe that MRAM improves the iso-area capacity by 2.8-3.2x while reducing the leakage power by a factor of 1.9-4.4. 
In the next section, we will evaluate the system-level implications of such MRAM write optimizations on the hardware performance of DNN training accelerators, where both area constraints and off-chip DRAM accesses are taken into account.

\vspace*{10pt}

\subsection{System-level analysis of energy efficiency}
In this subsection, we are going to discuss the system level results of STT-MRAM in comparison with SRAM as an on-chip scratchpad for a suite of DNN benchmarks. The DNNs that we chose for our evaluation are from image recognition tasks- AlexNet, VGG16, ResNet18 and ResNet34. The data sets used are CIFAR10 and CIFAR100. A batchsize of 512 is considered for the evaluation. There are two different scenarios used for the system level evaluation- iso-capacity and iso-area. Iso-capacity signifies the same capacity being used for SRAM and STT-MRAM memory arrays. Iso-area signifies same area being used for both STT-MRAM and SRAM arrays with different memory capacities. As mentioned in Section \ref{sec:exptsetup}, an output stationary dataflow for a systolic array architecture was used for the evaluation. Subsequently, we will discuss about the results for the two different scenarios. We have used VGG16 as the benchmark to study the two scenarios in details. 
\subsubsection{Iso-capacity}
\begin{figure}[htb]
  \centering
  \includegraphics[width=\columnwidth]{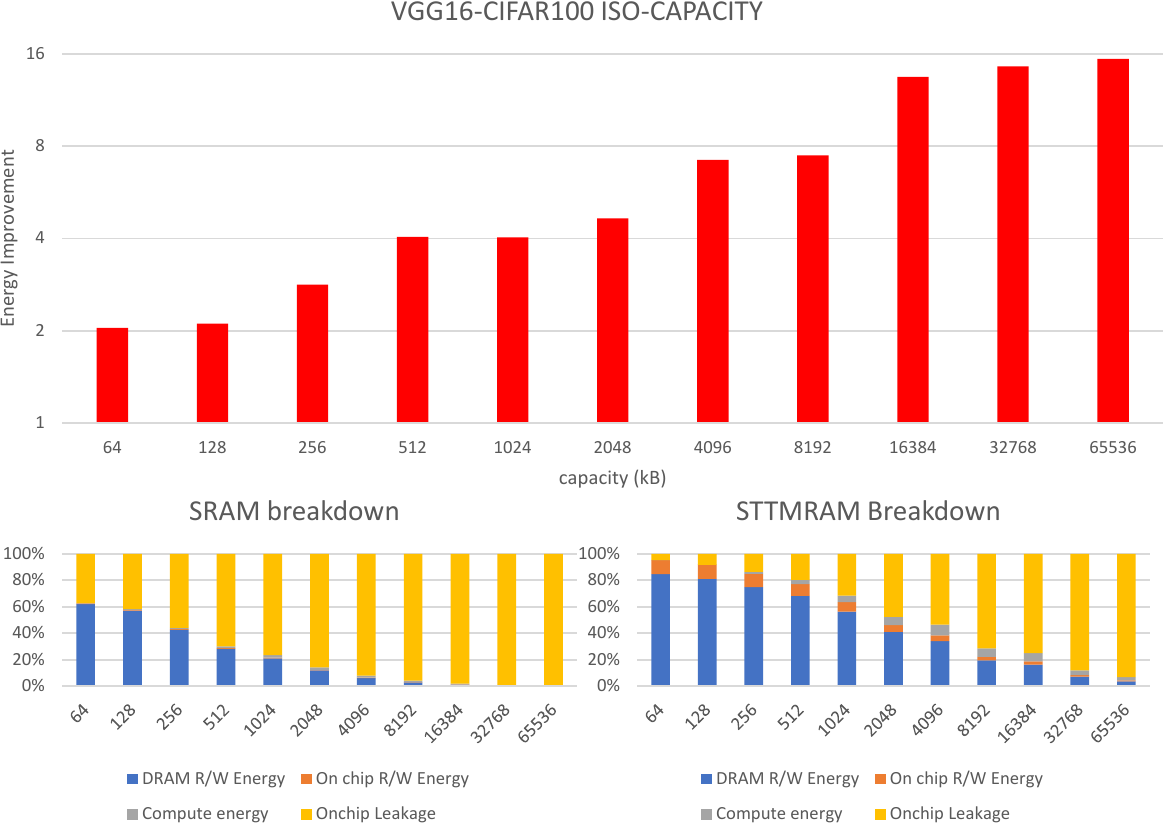}
  \caption{Iso capacity energy improvements of STTMRAM over SRAM on chip array for VGG16 evaluated on CIFAR100 and sytem energy breakdown for different components while using SRAM and STTMRAM as an on-chip scratchpad}
  \vspace{-4mm}
  \label{fig:vggisocapacity}
\end{figure}
\begin{figure}[htb]
  \centering
  \includegraphics[width=\columnwidth]{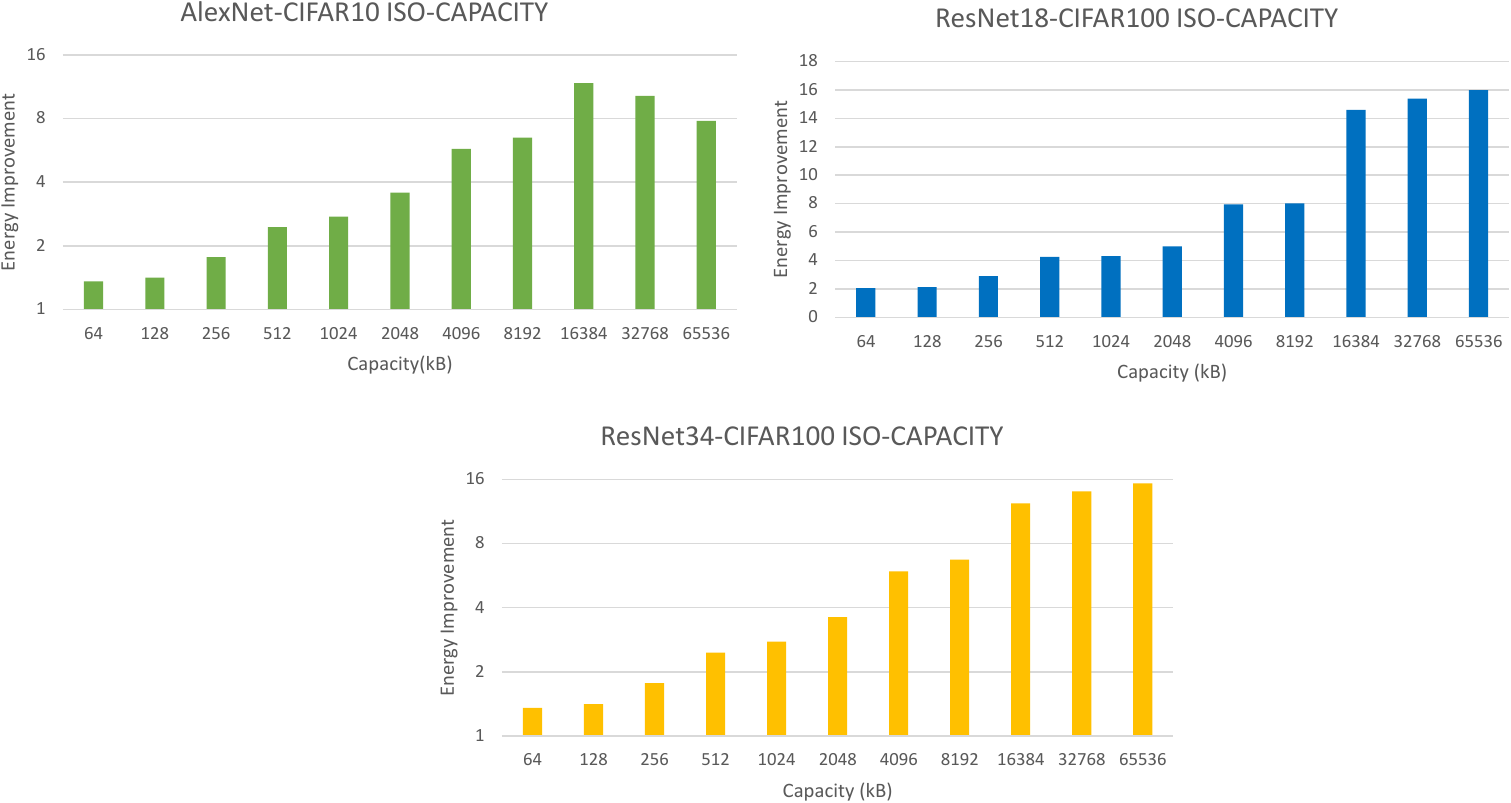}
  \caption{Iso capacity energy improvements of STTMRAM over SRAM on chip array for AlexNet, ResNet18 and ResNet34 evaluated on CIFAR10 and CIFAR100}
  \vspace{-4mm}
  \label{fig:allisocapacity}
\end{figure}
In this particular scenario, both SRAM and STTMRAM on-chip memory arrays maintain the same capacity. Fig \ref{fig:vggisocapacity} illustrates the energy improvement of STT-MRAM compared to SRAM for VGG16 on CIFAR100 at iso capacities with varying memory capacities. STT-MRAM exhibits a remarkable energy improvement of up to 15x over SRAM at iso capacity for different memory capacities in the case of VGG16 evaluated on CIFAR100. In this scenario, both DRAM accesses and on-chip access remain unchanged. However, the energy difference arises from the on-chip read and write energy, as well as the on-chip leakage energy. The improvement primarily stems from the reduction in leakage energy. The enhancements are more pronounced at higher memory capacities, mainly due to the diminishing gap in write energy between SRAM and STTMRAM. The breakdown of the different energy components can be observed in Fig \ref{fig:vggisocapacity}. For smaller capacities, the major contributors to the total system energy are DRAM access energy and on-chip leakage energy. On-chip accesses account for a smaller portion of the total energy. The compute energy percentage is negligible when compared to other energy components. As the memory capacity increases, the number of DRAM accesses decreases, and on-chip leakage becomes the dominant component of the total energy. This trend holds true for both SRAM and STTMRAM. However, due to the lower leakage power of STTMRAM arrays compared to SRAM arrays, the percentage of leakage in the overall energy is slightly lower for STTMRAM at iso capacity. STT-MRAM arrays have a higher contribution from on-chip access energy to the total energy compared to SRAM arrays. This is primarily due to the higher write energy of STTMRAM arrays in comparison to SRAM. The graphs in Fig \ref{fig:allisocapacity} demonstrate the energy improvement of STTMRAM over SRAM for AlexNet, ResNet18, and ResNet34 evaluated on CIFAR10, and CIFAR100 at iso capacities with different memory capacities. Similar trends can be observed across all benchmarks, akin to the case of VGG16. The improvement remains consistent across different capacities, with smaller improvements at lower capacities and more significant improvements as capacity increases, primarily due to the widening gap in leakage and the closing gap between write energy of SRAM and STTMRAM.
\subsubsection{Iso-area}
\begin{figure}[htb]
  \centering
  \includegraphics[width=\columnwidth]{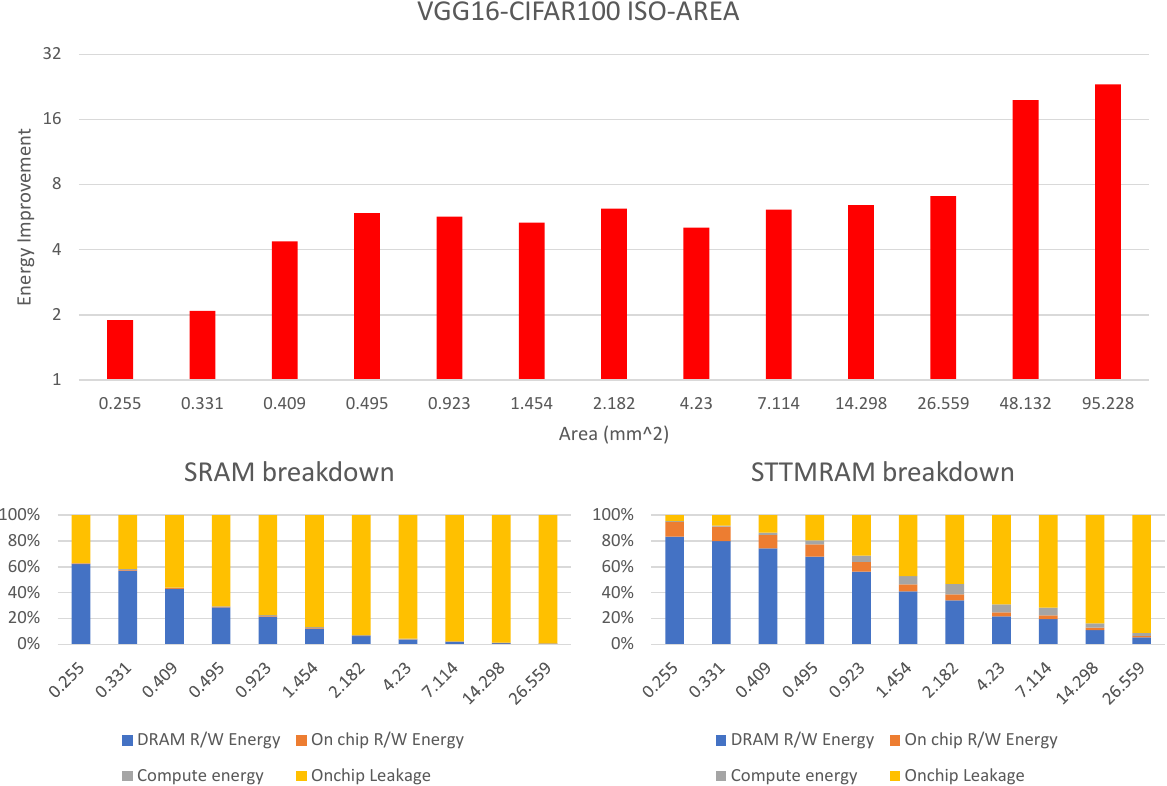}
  \caption{Iso area energy improvements of STTMRAM over SRAM on chip array for VGG16 evaluated on CIFAR100 and sytem energy breakdown for different components while using SRAM and STTMRAM as an on-chip scratchpad}
  \vspace{-4mm}
  \label{fig:vggisoarea}
\end{figure}

\begin{figure}[htb]
  \centering
  \includegraphics[width=\columnwidth]{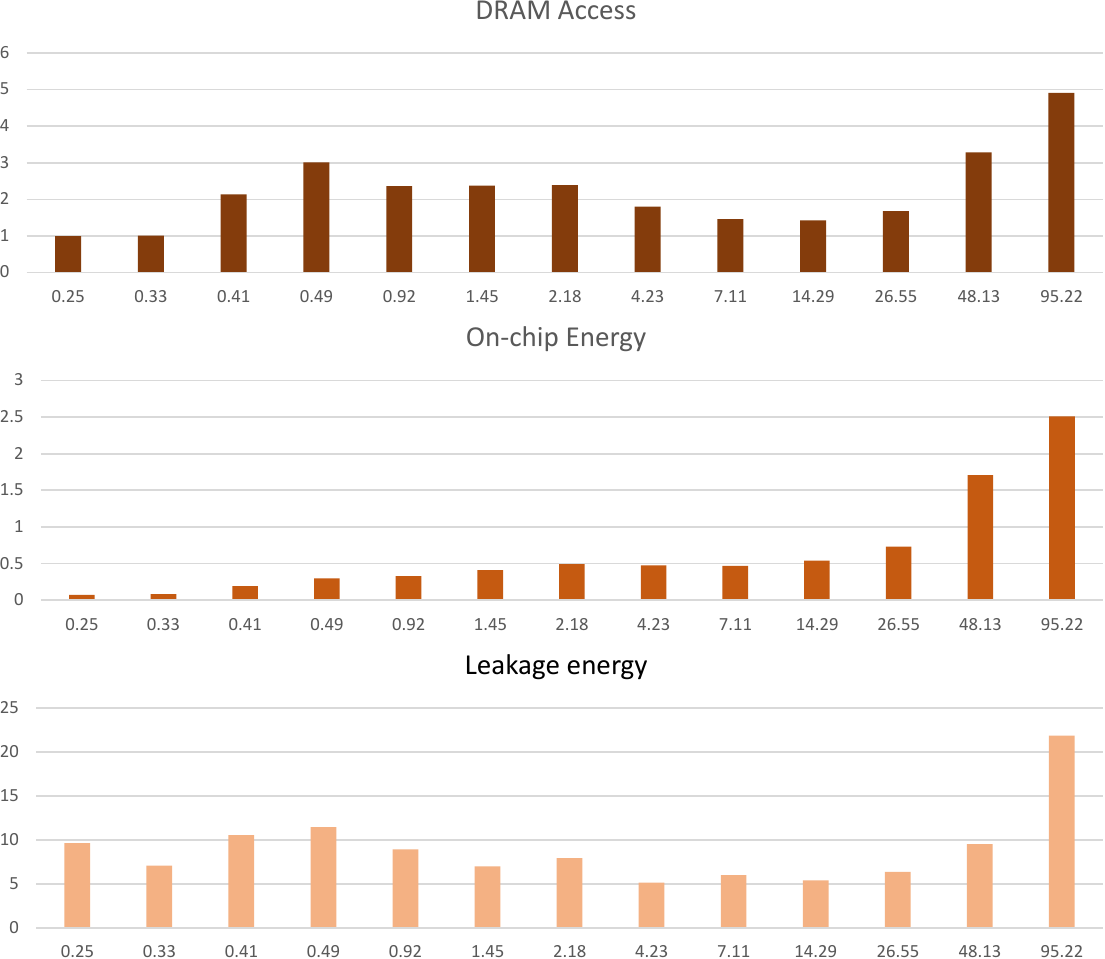}
  \caption{Iso area energy ratio (STTMRAM over SRAM) of different energy components for VGG16 evaluated on CIFAR100}
  \vspace{-4mm}
  \label{fig:vggisoareabreakdown}
\end{figure}

\begin{figure}[htb]
  \centering
  \includegraphics[width=\columnwidth]{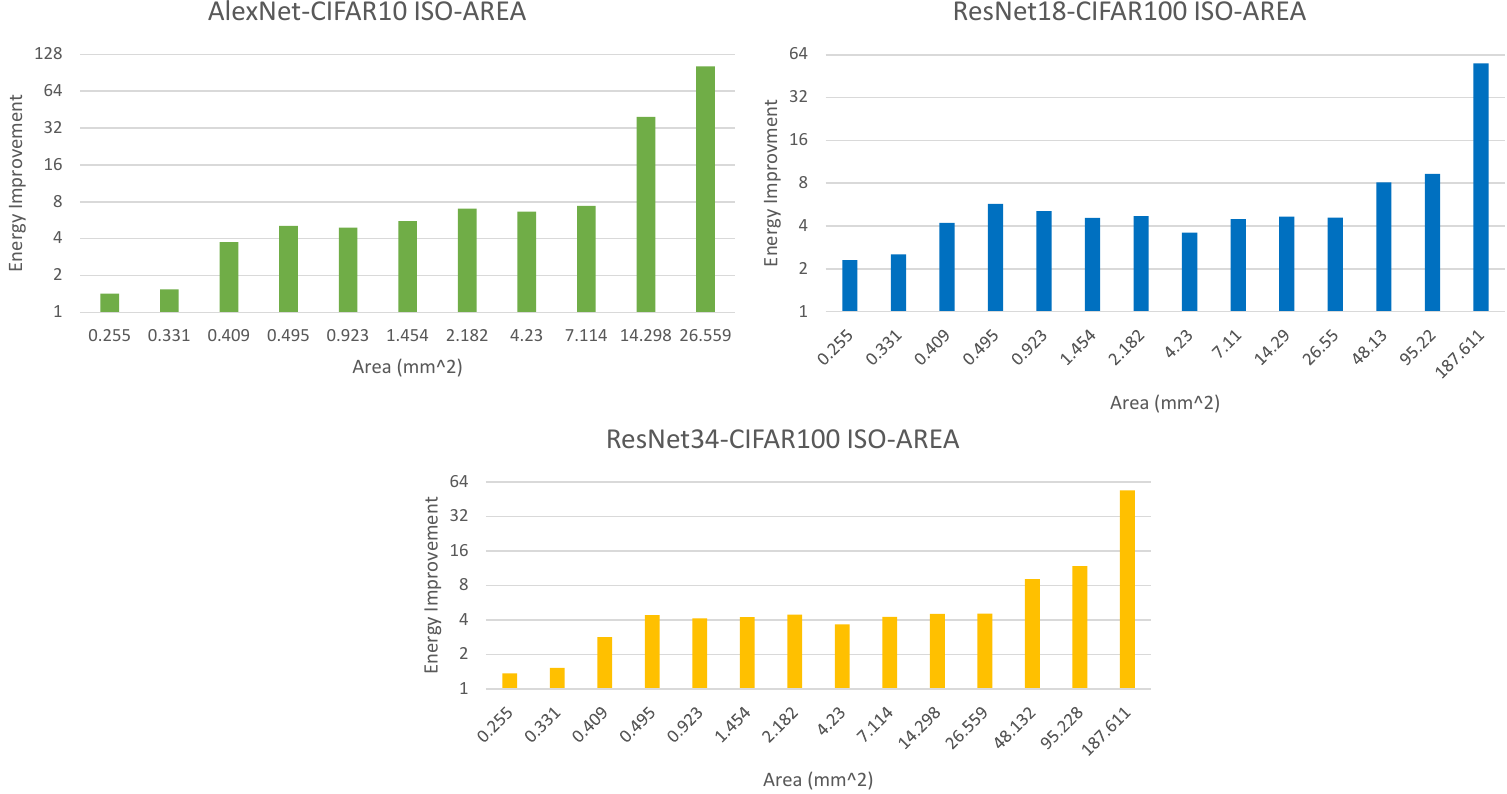}
  \caption{Iso area energy improvements of STTMRAM over SRAM on chip array for AlexNet, ResNet18 and ResNet34 evaluated on CIFAR10 and CIFAR100}
  \vspace{-4mm}
  \label{fig:allarea}
\end{figure}

In this scenario, we maintain the same on-chip area for both SRAM and STT-MRAM memory arrays. We adjust the SRAM memory capacities while keeping the capacity for STT-MRAM the same, ensuring the on-chip area remains unchanged. Figure \ref{fig:vggisoarea} illustrates the energy improvement of STT-MRAM over SRAM when used as an on-chip scratchpad at iso area, considering different area values for VGG16 evaluated on CIFAR100. At a specific capacity, STT-MRAM requires fewer DRAM accesses compared to SRAM memory arrays. Additionally, the on-chip accesses differ, with variations in energy stemming from on-chip read and write operations, on-chip leakage, and DRAM access. The ratio of these energy components for SRAM over STT-MRAM is depicted in Figure \ref{fig:vggisoareabreakdown}. As expected, the energy consumption of DRAM is higher for SRAM than for STT-MRAM when utilized as a scratchpad across various area values. The improvement in the DRAM access energy ratio fluctuates due to the varying capacity ratios between SRAM and STT-MRAM. The increased on-chip capacity of STT-MRAM arrays compared to SRAM arrays accounts for the enhancements in DRAM access. The number of on-chip accesses depends on the capacity, and variations in replacement policy and capacity lead to different numbers of on-chip accesses for different area values. As anticipated, SRAM generally exhibits lower on-chip energy than STT-MRAM for most areas due to significantly lower write energy. At higher area values, it is possible for on-chip accesses to be lower for STT-MRAM than for SRAM, resulting in a slight improvement in on-chip access energy. The memory array results indicate that the gap between SRAM and STT-MRAM write energy decreases at higher capacity values. Even with different on-chip capacities, we can expect the write energy of SRAM and STT-MRAM to be closer at higher memory capacities (larger areas). In scenarios where on-chip accesses are smaller for STT-MRAM arrays compared to SRAM arrays, it is possible for STT-MRAM arrays to have lower overall write energy. Notably, there is a significant improvement in leakage energy, albeit lower than in the iso-capacity case, due to the higher capacity of STT-MRAM compared to SRAM for a specific area. The trend of increasing leakage energy with increasing area is not as clear as in the iso-scenario case since the capacity ratios differ across different area values. Overall, at iso area, STT-MRAM memory arrays offer significant benefits over SRAM memory arrays for all area values. The energy improvement of STT-MRAM over SRAM for AlexNet, ResNet18, and ResNet34 evaluated on CIFAR10, and CIFAR100 at iso area for various area values is depicted in the graphs of Figure \ref{fig:allarea}. Similar trends to the VGG16 case are observed for all other benchmarks. In the case of AlexNet, there is a sudden increase in energy improvement. At a certain area value, the capacity of STT-MRAM becomes large enough to accommodate all the required data structures for training, whereas the SRAM memory array cannot due to its lower capacity. Hence, we observe a sudden spike in energy improvement, which we do not include in the reported results since it is only observed at a narrow range of areas. As the area value increases further, the SRAM capacity also becomes capable of accommodating all the data structures, leading to a drop in energy improvement, although there is still a significant improvement for STT-MRAM. If we increase the area values in the other benchmarks, a similar trend can be expected, although not captured in the current graphs. Therefore, the trend heavily depends on the network type and dimensions.

\subsection{Impact on Write Optimization on Application Level Accuracy and Write Energy Benefits}
\begin{figure}[htb]
\begin{subfigure}
  \centering
  \includegraphics[width=\linewidth]{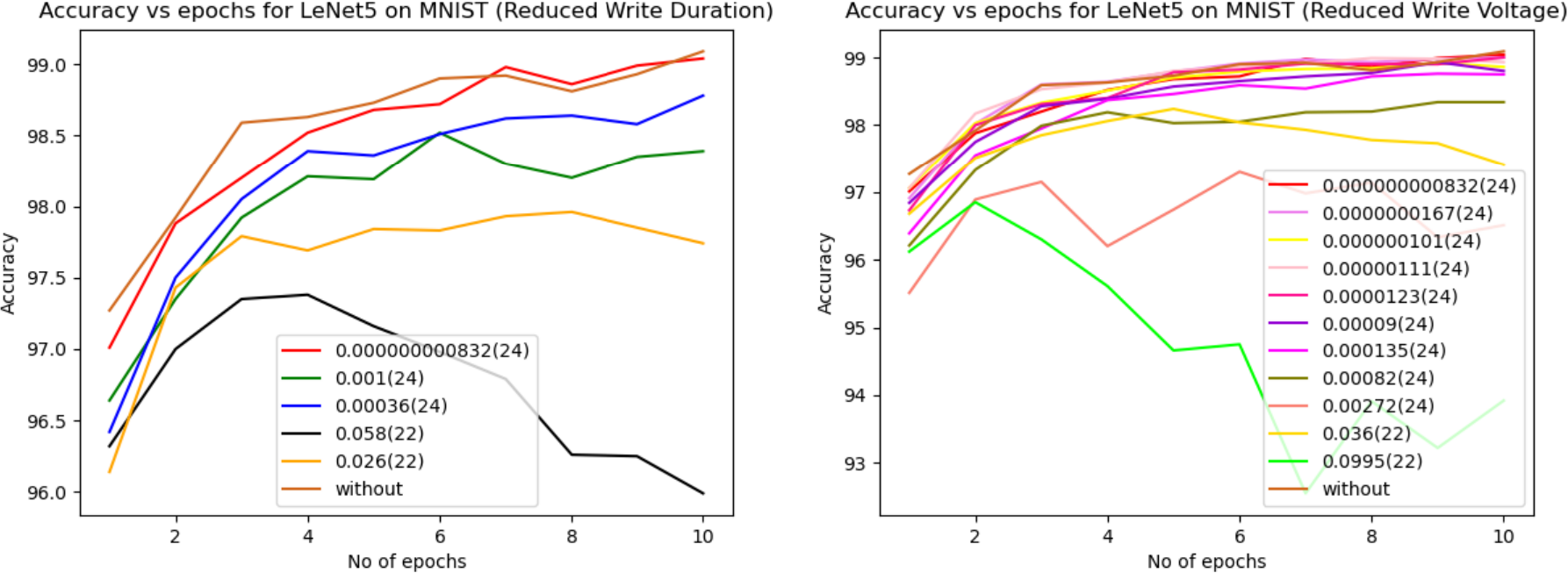}
  
\end{subfigure}
\begin{subfigure}
  \centering
  \includegraphics[width=\linewidth]{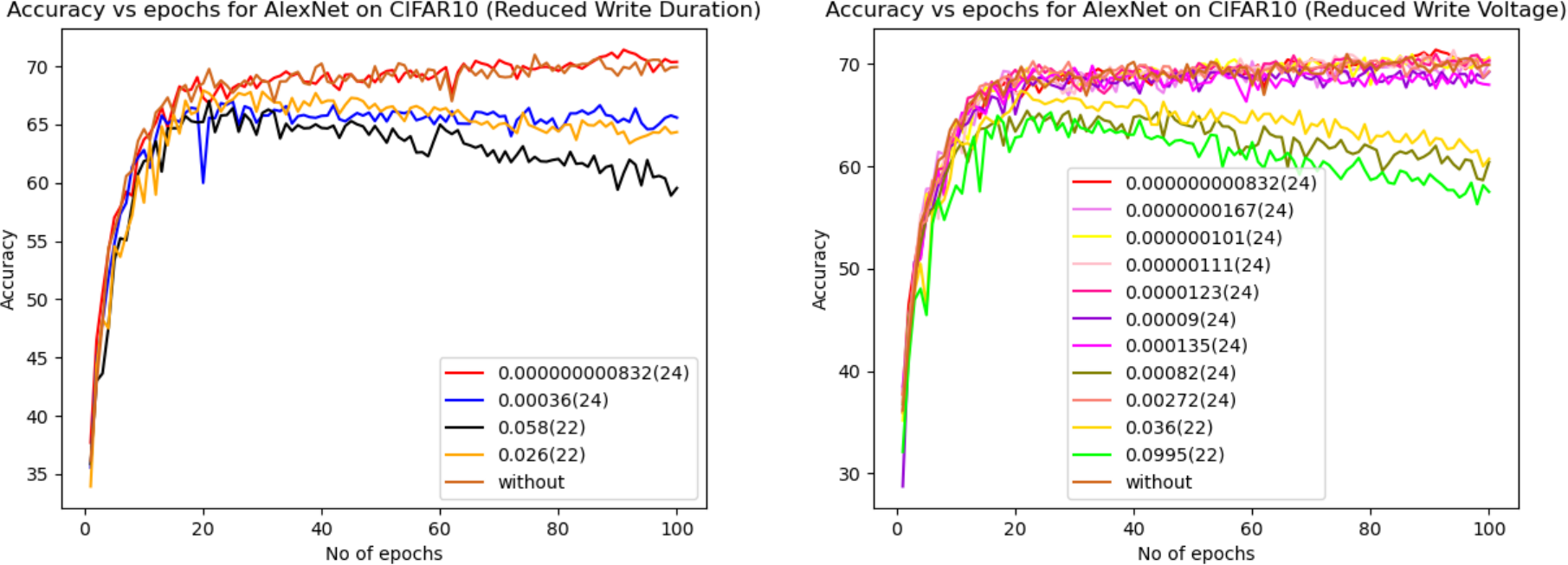}
\end{subfigure}
\begin{subfigure}
  \centering
  \includegraphics[width=\linewidth]{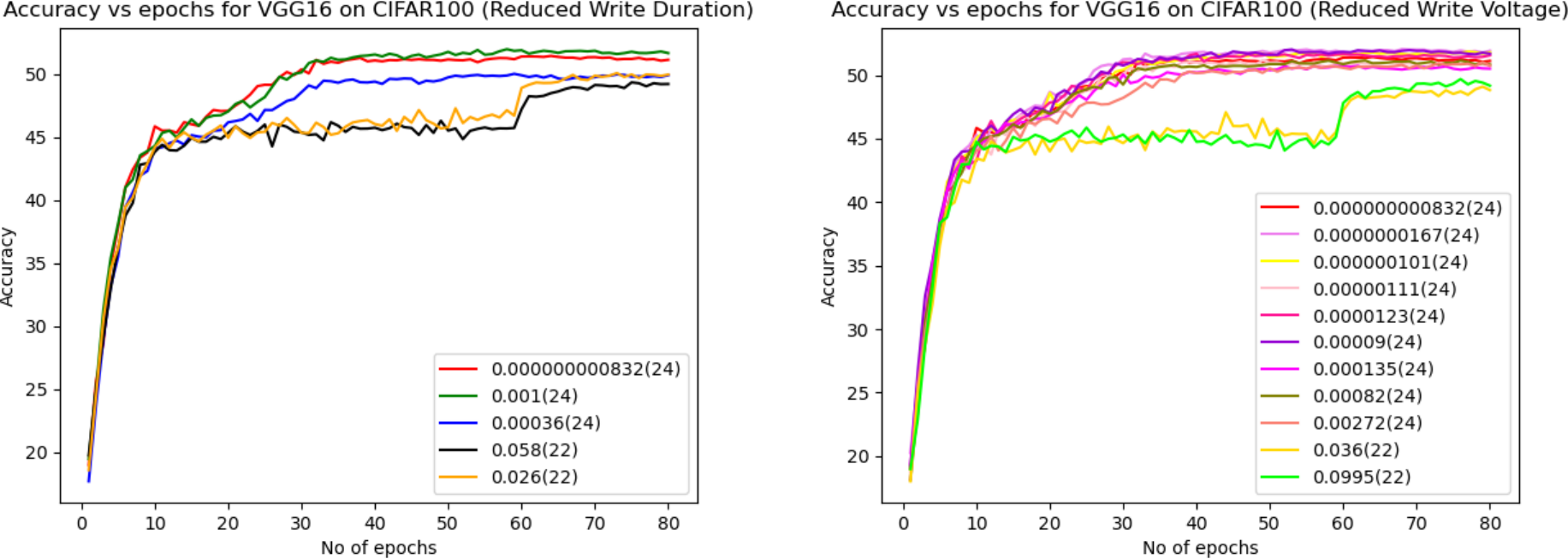}
\end{subfigure}
\begin{subfigure}
  \centering
  \includegraphics[width=\linewidth]{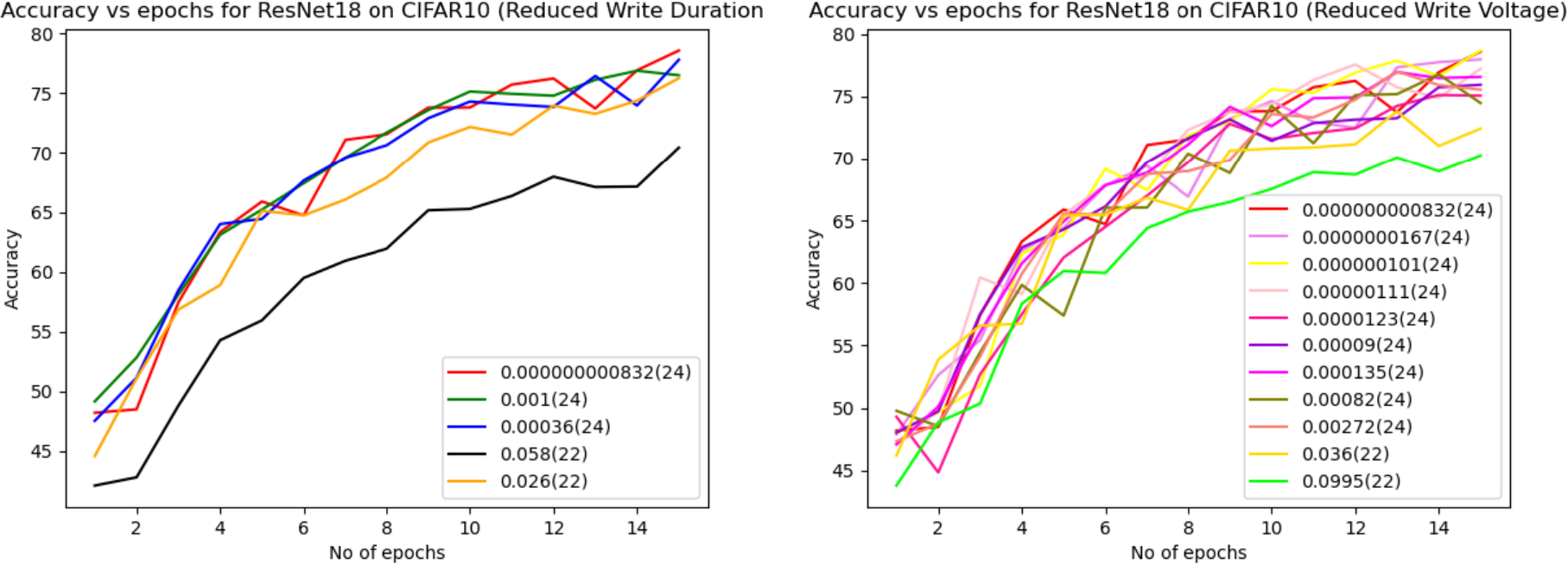}
\end{subfigure}
\caption{Application level accuracy plots for different write error rates as a result of STT-MRAM write optimizations for LeNet, AlexNet, VGG16 and ResNet18 evaluated on MNIST, CIFAR10 and CIFAR100}
\label{fig:error_accuracy}
\end{figure}
\begin{figure}[htb]
  \centering
  \includegraphics[width=\columnwidth]{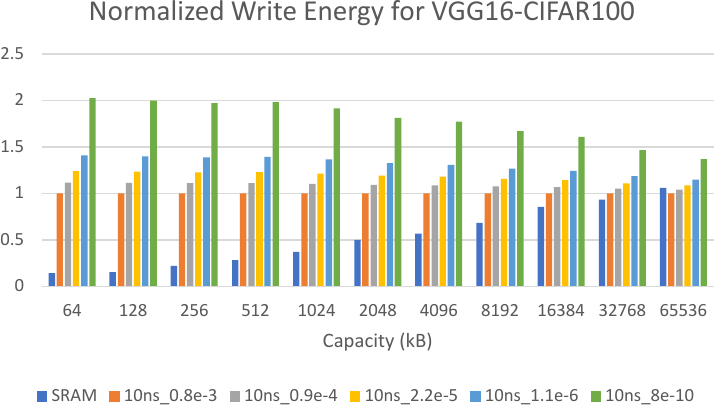}
  \caption{Normalized write energy for VGG16 evaluated on CIFAR100 for iso-capacity scenario. The write energy is based on using the optimized STT-MRAM for only the mantissa bits and the base STT-MRAM for the exponent and sign bits}
  \vspace{-4mm}
  \label{fig:writenergy}
\end{figure}
In previous subsections, we observed that the write energy of STT-MRAM arrays was significantly higher than that of SRAM. However, it is possible to optimize STT-MRAM writes by reducing the voltage and duration of the write operation, albeit at the expense of write errors. In this subsection, we evaluate the impact of these write optimizations on application-level accuracy and the corresponding system-level write energy benefits. To assess the effect of write errors on application-level accuracy, errors were introduced at the bit level in single precision floating-point data format. We considered the IEEE754 data format, which consists of 1 sign bit, 8 exponent bits, and 23 mantissa bits. We emulated error-based training in PyTorch for some of the aforementioned benchmarks. Our results indicated that significant bits (1 sign bit and 8 exponent bits) can cause significant variations in gradient values for high write error rates, leading to error propagation and non-convergence of the training process. On the other hand, higher significant bits can tolerate small error rates. Therefore, we decided to apply errors to the least significant bits of the floating-point data format (mantissa bits). The training results for different error rates are presented in Figure \ref{fig:error_accuracy} for LeNet5, AlexNet, VGG16, and ResNet18 evaluated on MNIST, CIFAR10, and CIFAR100. The legends in the graph indicate the write error rate, and the numbers in brackets signify the number of bits from the floating-point data format considered for introducing errors based on the error probability. Training with errors at the bit level is time-consuming even on a GPU, so for some benchmarks, only a few epochs are shown. It is evident from Figure \ref{fig:error_accuracy} that partitioning words among different arrays with varying levels of reliability can lead to minimal accuracy degradation in training, even for higher write error rates. Lower error rates show almost no drop in application-level accuracy. We attempted to evaluate the system-level write energy by utilizing a heterogeneous memory array, where the least significant bits (mantissa bits) are mapped to the optimized STT-MRAM array, while the higher significant bits (exponent and sign) are mapped to the base STT-MRAM array. Figure \ref{fig:writenergy} illustrates the normalized system-level write energy for the heterogeneous memory array at different error rates for different memory capacities. The energy numbers are normalized to the STT-MRAM array with the highest error rate (0.8e-3). We also included the normalized SRAM number for comparison. In this optimization, the write latency of STT-MRAM is fixed at 10ns, and the write voltage is varied to generate the different error rates shown in the graph. It is evident that higher error rates bring the system-level write energy closer to the write energy of SRAM. Tolerating a high error rate can result in up to a 2.03x improvement in total write energy compared to unoptimized STT-MRAM for VGG16 evaluated on CIFAR100. Hence, selectively choosing the bits for the optimized STT-MRAM array can provide a significant improvement in system-level write energy with minimal application-level accuracy degradation.


\vspace*{6pt}
\section{Conclusion}
\label{sec:conclusion}
{\noindent} 
 
We demonstrate a device-to-algorithm co-exploration towards efficient implementation of embedded STT-MRAM as the scratchpad for ML training acceleration. We exploit the error-resiliency of DNNs and utilize low-cost MRAM write operations with reduced voltage and pulse duration. In order to mitigate the elevated write errors due to the probabilistic switching dynamics of MRAM, we propose heterogeneous memory architecture where the different bit segments (sign/exponent/mantissa) of input and weight parameters are 
mapped into MRAM sub-arrays with distinct write configurations. Our device-to-system simulation shows that by replacing the standard CMOS-based SRAM with STT-MRAM, substantial energy improvement is achieved due to the improved density and reduction of leakage power. Moreover, our cross-layer evaluation demonstrate that MRAM with reduced write costs can further harness improvement in the memory write energy while maintaining the training convergence and prediction accuracy of DNNs. Our proposed cross-layer co-optimization essentially closes the performance gap  between MRAM and SRAM (in terms of write latency and energy consumption) at the array-level, leading to a promising pathway of designing high-performing ML training accelerators with high-density and energy-efficient NVMs.

\section{Acknowledgment}
This work was supported by C-BRIC, one of six centers in JUMP, a Semiconductor Research Corporation (SRC) program, sponsored by DARPA.
\vspace*{-0pt}
\scriptsize
\bibliographystyle{unsrt}
\bibliography{paper}

\end{document}